\documentclass[11pt,twoside]{article}

\usepackage{asp2006}
\usepackage{epsf}
\usepackage{psfig}
\usepackage{lscape}
\usepackage{rotating}

\begin{document}
\title{Population synthesis models including AGB stars and their ingredients} 
\author{Paola Marigo}
\affil{}{Dipartimento di Astronomia, Universit\`a di Padova, Italy}

\begin{abstract} 
I will briefly review the state of the art of evolutionary 
population synthesis (EPS) 
models that include the contribution from AGB stars.
\end{abstract}

\section{Assembling AGB stars in EPS models: techniques}
\label{sec_tech}
Three possible methods can be distinguished 
to add the AGB phase in EPS models,  which are based on the
use of A) stellar isochrones (e.g. Bressan et al. 1996, 
Bruzual \& Charlot 2003, Mouhcine 2002, Marigo et al. 2003); B)  
stellar tracks (Groenewegen \& de Jong 1993, Marigo et al. 1999, 
Mouhcine \& Lan\c{c}on 2002ab, Mouhcine \& Lan\c{c}on 2003); 
and C) the fuel consumption theorem by Renzini \& Buzzoni (1986; 
Maraston 1998, 2005).

Both methods A) and B) need essentially the same ingredients, namely 
sets of AGB stellar models that provide the 
evolution of basic stellar parameters (i.e. total mass, core mass, luminosity,
effective temperature, surface chemical composition) 
with a possibly wide coverage of initial stellar masses and metallicities.  
Owing to the heavy requirement of computing time set by the full 
modelling of the AGB phase, the most convenient way to 
build up an extended library of AGB stellar tracks is to adopt
a synthetic approach, 
that gives a simplified but flexible and easy-to-test
description of the stellar evolution on the AGB. Another positive aspect is
that synthetic AGB models can be readily 
calibrated on the base of basic observables
(e.g. the carbon star luminosity functions), hence ensuring a more realistic
description of AGB properties (see Sect.~\ref{sec_obscons}).
Available stellar isochrones including the whole AGB phase are listed in 
Table~\ref{tab_isoc}, that also outlines a few relevant characteristics,  
namely: the inclusion or not of the third dredge-up and hot-bottom burning, 
the way $T_{\rm eff}$ is obtained depending on the adopted
molecular opacities, the consideration or not of 
the effect of circumstellar dust on the emitted spectrum.   

The fundamental ingredient of Method C) is the fuel matrix, 
$F_{\rm AGB}(M_{\rm TO}, Z)$, providing the nuclear fuel burnt during the
AGB phase as a function of the turn-off mass
$M_{\rm TO}$ (or equivalently of the age), 
and the metallicity $Z$ of the corresponding single stellar population. 

Then, for all methods the passage from theoretical quantities 
to observed ones requires the adoption of suitable  spectral libraries
as a function of $T_{\rm eff}$, surface gravity and C/O ratio,
or alternatively tables of bolometric corrections and 
$T_{\rm eff}$--colour transformations. Finally, one has 
to specify global properties of the simulated galaxy, 
like the initial mass function (IMF), the star
formation rate history (SFR), and the age-metallicity relation (AMR). 

Concerning the characteristics of the EPS techniques, it should be remarked 
that both methods A) and B) are 
tied, by construction, to the adopted stellar models.
On one hand  this brings along all the weak points 
that affect the theory of stellar evolution, 
but on the other hand it allows to get a useful feedback
to input prescriptions any time a comparison between predictions and
observations is performed.   
Method C) usually adopts an empirical calibration of the AGB nuclear fuel based
on observed data of Magellanic Clouds' clusters. On one hand this  
guarantees to account for the correct integrated 
light contribution from AGB stars at LMC and SMC metallicities,
but on the other hand it makes it difficult to derive explicit indications  
on uncertain aspects of the underlying AGB evolution.
Moreover, while for methods A) and B) the resolution element is the single
star, so that they can address the study of both integrated and resolved
stellar populations, for method C) the maximum resolution is given
by the  simple stellar population (SSP), 
which limits its application to the integrated properties 
of unresolved AGB stars.

Because of their intrinsic sensitivity to the details of AGB modelling,
this review will mostly focus  
on EPS methods A) and B) and their applications.

\section{Synthetic AGB evolution: ingredients}
\label{sec_synagb}
Synthetic AGB models are usually constructed on the base of 
some analytical recipe -- derived from complete modelling of the AGB phase
and/or observations --  providing,  for instance, the core mass-luminosity
relation, the core mass-interpulse period relation, and the mass-loss rates
as a function of stellar parameters.  Examples of such calculations can be
found in Groenewegen \& de Jong (1993) and Izzard et al. (2004).

The purely analytical scheme can be importantly complemented with the
aid of envelope integrations - from the photosphere down to the surface
of the H-exhausted core -- 
which allows us e.g. 
to predict the effective temperature along the AGB, as well
as to follow the hot-bottom burning (HBB) nucleosynthesis 
by solving a network of nuclear reactions
in the deepest envelope layers  of the most massive AGB models.
This kind of approach was pioneered by Renzini \& Voli (1981), and 
more recently adopted by e.g. Marigo et al. (1999 and references therein), 
Gavil\'an et al. (2005), Marigo \& Girardi (2006). 

\begin{sidewaystable}
\small
\caption{Available stellar isochrones including the AGB phase}
\begin{tabular}{lcccl}
\hline
Reference$^{(\rm a)}$ & $3^{\rm rd}$D-up and HBB$^{(\rm b)}$ & $T_{\rm eff}$$^{(\rm c)}$ & Dust$^{(\rm d)}$ & available at$^{(\rm e)}$ \\
\hline
Bertelli et al. 1994, A\&AS, 106, 275    &  NO  &  solar-scaled comp.  &  NO  &   http://pleiadi.pd.astro.it \\
Girardi et al. 2000, A\&AS, 141, 371  &     NO  &  solar-scaled  comp. &  NO    &        http://pleiadi.pd.astro.it\\
Marigo \& Girardi 2001, A\&A, 377, 132   &   YES &  solar-scaled  comp. &  NO   &         http://pleiadi.pd.astro.it \\
Pietrinferni et al. 2004, ApJ, 612, 168    &  NO & solar scaled comp.  &   NO    & http://astro.ensc-rennes.fr/basti \\
Cioni et al. 2006,  A\&A, 448, 77  & YES  & O-/ C-rich comp. &    NO  &          http://pleiadi.pd.astro.it  \\
Bressan et al. 1998, A\&A, 332, 135  & NO  &  solar scaled comp. &   YES &               upon request \\
Mouhcine 2002, A\&A, 394, 125  & YES    &  solar scaled comp. &    YES  &               upon request \\
Piovan et al. 2003, A\&A, 408, 559 &  NO  &  solar scaled comp. &   YES  &                upon request \\
Kim et al. 2006, PASP, 118, 62  & NO   &  solar scaled comp.  &     YES &                 upon request\\
\hline
\multicolumn{5}{l}{\footnotesize{$^{(\rm a)}$: bibliographic reference}}\\
\multicolumn{5}{l}{\footnotesize{$^{(\rm b)}$: 
explicit inclusion or not of AGB nucleosynthesis, 
i.e. third dredge-up and hot-bottom burning}, in the underlying AGB models} \\
\multicolumn{5}{l}{\footnotesize{$^{(\rm c)}$: choice of 
low-temperature opacities affecting the predicted effective 
temperature,}}\\ 
\multicolumn{5}{l}{\hspace{0.58cm}\footnotesize{i.e. for solar-scaled or 
variable (O- or C-rich) chemical mixtures}} \\
\multicolumn{5}{l}{\footnotesize{$^{(\rm d)}$: inclusion or not 
of circumstellar dust in the computation of the emitted spectrum}} \\
\multicolumn{5}{l}{\footnotesize{$^{(\rm e)}$: 
public www address for retrieval }}\\
\hline
\end{tabular}
\label{tab_isoc}
\vspace{1.cm}
%
\caption{Available chemical yields from low- and intermediate-mass stars}
\begin{tabular}{lcccll}
\hline
Reference$^{(\rm a)}$  &  model type$^{(\rm b)}$  &  HBB$^{(\rm c)}$  &   isotopes$^{(\rm d)}$ & \multicolumn{1}{c}{$M/M_{\odot}$$^{(\rm e)}$} &  \multicolumn{1}{c}{$Z$$^{(\rm f)}$}\\
\hline
Iben \& Truran 1978, ApJ, 220, 980 &  synthetic &  NO  &  C, N, $^{22}$Ne &  $1.0-8.0$ & $0.02$ \\
Renzini \& Voli 1981, A\&A, 94, 175  & synthetic+env. &  network  &  He  C, N, O & $1.0-8.0 $  &  $0.004-0.02$ \\
v.d. Hoek \& Groenewegen 1997, A\&AS, 123, 305 &   synthetic &   analytic &  He  C, N, O &      $0.8-8.0 $  & $0.001-0.04$ \\
Forestini \& Charbonnel  1997, A\&AS, 123, 241  &  synthetic/full &  network/extr. &   many &   $3.0-7.0 $    & $0.005-0.02$ \\
Marigo  2001, A\&A, 370, 194  &  synthetic+env. &  network  & He  C, N, O & $0.8-5.0 $    & $0.004-0.019$ \\
 Chieffi et al. 2001, ApJ, 554, 1159  &  full   & network & many  & $4.0-7.0 $  & $0$ \\
Ventura et al. 2002, A\&A, 393, 215 & full & network &  He, Li,  C, N, O & $2.5-5.5 $   & $2\: 10^{-4} - 0.01$ \\
Karakas 2003, PhD thesis, Monasch Univ. & full  & network &   many  & $1.0-8.0 $    & $0.004- 0.02$ \\
Dray et al. 2003, MNRAS, 338, 973   &  full   &  network &  C, N, O  & $1.25-6.0 $   & $0.02$ \\
Herwig 2004, ApJS, 155, 651  &  full   &  network & He, C, N & $2.0-6.0$  & $10^{-4}$ \\
 Izzard et  al. 2004, MNRAS, 350, 407 &  synthetic &  analytic &  He, C, N, O,  $^{22}$Ne &     $0.8-8.0 $     & $10^{-4}-0.03$ \\
Gavil\'an et al. 2005,  A\&A, 432, 861    & synthetic+env. &  network & He, C, N, O  & $0.8-8.0  $   & $ 0.013-0.032$   \\
\hline
\multicolumn{6}{l}{\footnotesize{$^{(\rm a)}$: bibliographic reference}}\\
\multicolumn{6}{l}{\footnotesize{$^{(\rm b)}$: kind of AGB model, 
i.e. {\em synthetic} $\rightarrow$ purely analytic; 
{\em synthetic+env.}  $\rightarrow$ analytic with envelope integrations;}}\\
\multicolumn{6}{l}{\hspace{0.58cm}\footnotesize{{\em full}  $\rightarrow$ complete stellar 
structure calculations }} \\
\multicolumn{6}{l}{\footnotesize{$^{(\rm c)}$: treatment of HBB, 
i.e. {\em analytic} $\rightarrow$ based on analytic recipe; 
{\em network} $\rightarrow$ integration of a nuclear network;}}\\
\multicolumn{6}{l}{\hspace{0.58cm}\footnotesize{{\em extr.} $\rightarrow$ extrapolation of partial results up to the end of the AGB}}\\
\multicolumn{6}{l}{\footnotesize{$^{(\rm d)}$: elemental species and their 
main isotopes for which stellar yields are available}}\\ 
\multicolumn{6}{l}{\footnotesize{$^{(\rm e)}$ and $^{(\rm f)}$: 
initial mass and metallicity intervals}} \\
\hline
\end{tabular}
\label{tab_yields}
\end{sidewaystable}

\subsection{Luminosity and effective temperature}
\label{ssect_lt}
\begin{figure}[!ht]
\plottwo{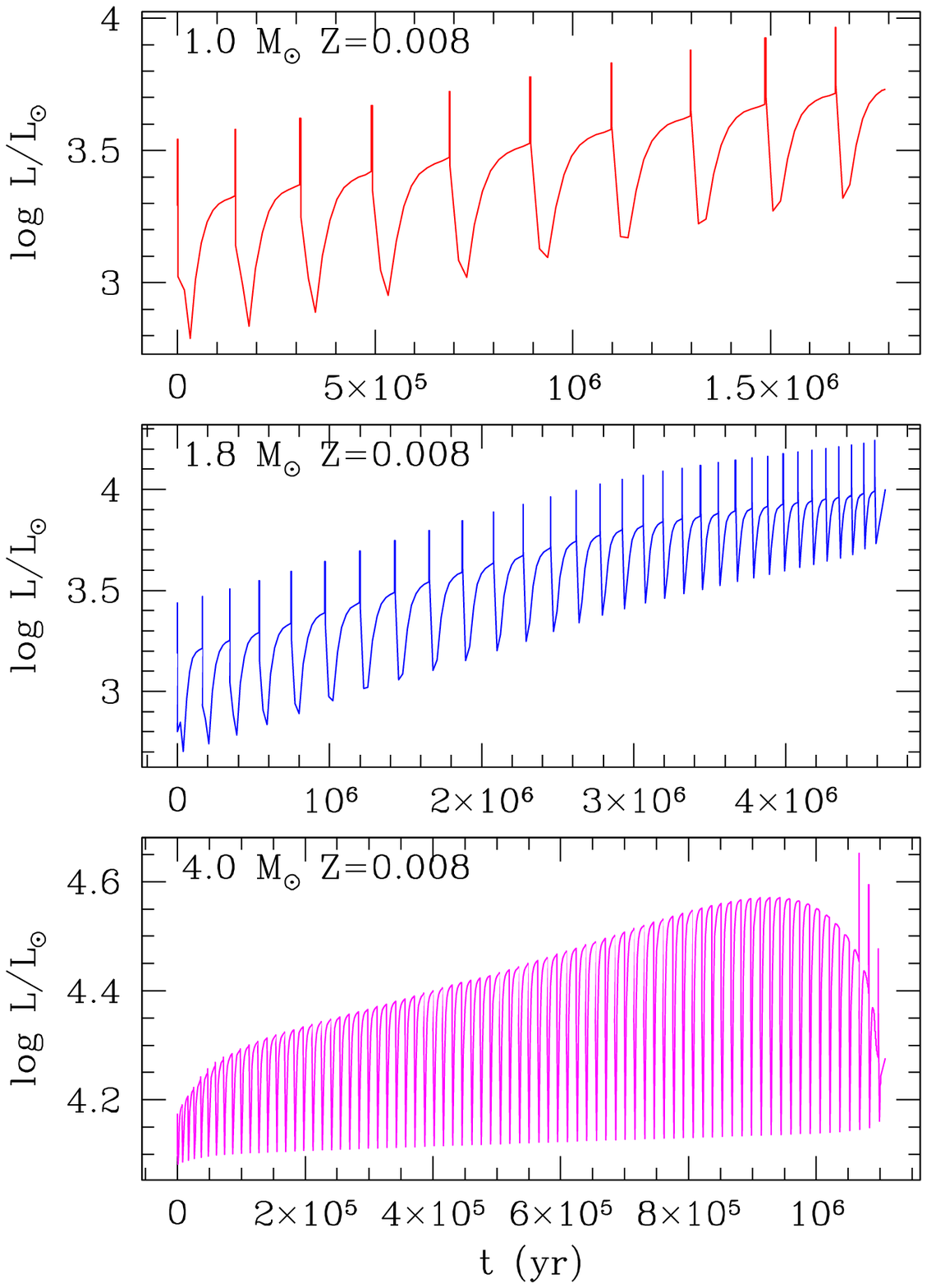}{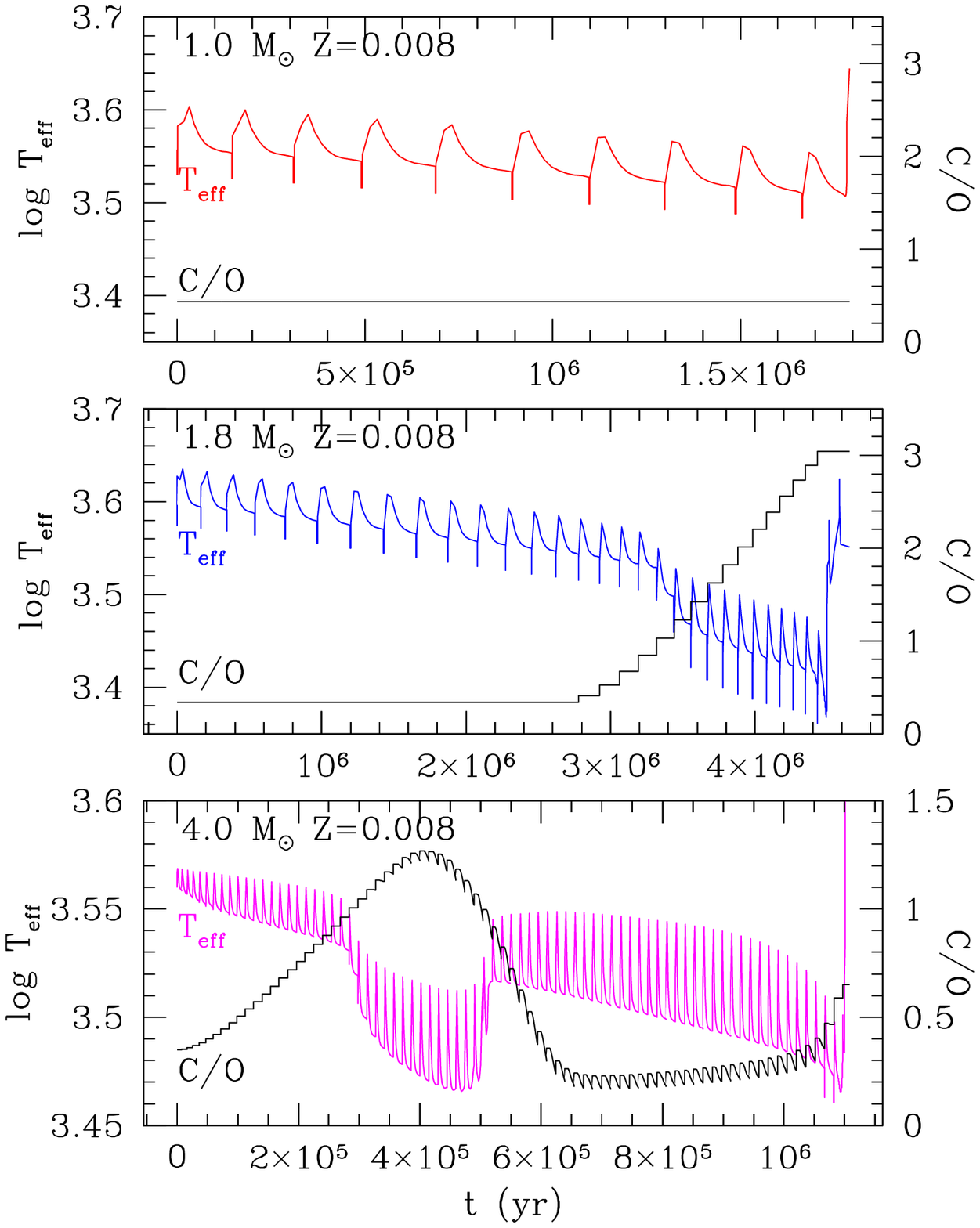}
\caption{Left panel: luminosity evolution for few selected synthetic TP-AGB 
models adopting the formalism developed by Wagenhuber \& Groenewegen (1998).
Right panel: Evolution of $T_{\rm eff}$ and surface C/O ratio 
for the same set of models,
obtained with the aid of envelope integrations including variable molecular opacities (Marigo 2002). 
Note the cooling effect as soon C/O overcomes unity due to the third
dredge-up.
In all cases calculations have been carried out from the first thermal pulse
up to the complete ejection of the envelope 
(Marigo \& Girardi 2006)}
\label{fig_lt}
\end{figure}

The recent availability  of high-accuracy formalisms based on full
evolutionary calculations (e.g. Wagenhuber \& Groenewegen 1998, 
Izzard et al. 2004) has allowed synthetic TP-AGB models to account 
for the complex behaviour of the luminosity due to the occurrence
of thermal pulses and HBB. An example of the nice performance of these 
formalisms is
presented in Fig.~\ref{fig_lt}, showing 
the flash-driven luminosity variations in models with different masses, and 
the over-luminosity effect due to HBB in the 
$(M_{\rm i}=4.0\,M_{\odot}, Z=0.008)$ model.

As to the effective temperature, several analytic fits to the results of
full AGB evolutionary calculations are available in the literature,
yielding $T_{\rm eff}=T_{\rm eff}(M,L,Z)$ (e.g. Vassiliadis \& Wood 1993). 
However, it should be mentioned that these fitting relations
are strictly valid for giant stars with C/O$<1$ as they stand on the 
adoption of low-temperature opacities for solar-scaled chemical compositions,
while they are completely inappropriate to describe the atmospheric properties
of carbon stars. 
As discussed by Marigo (2002), the adoption of molecular opacities 
consistently coupled to the surface chemical composition of AGB models leads
to  a significant decrease of $T_{\rm eff}$ as soon as   C/O$>1$, which
is confirmed by observations of galactic AGB stars (Bergeat et al. 2001).
The aforementioned  effect is clearly shown in the right panel of
 Fig.~\ref{fig_lt}, where the behaviour of $T_{\rm eff}$ appears 
to mirror somehow that of C/O.

\subsection{The third dredge-up and hot-bottom burning}
\label{ssec_dredhbb}

\begin{figure}[!ht]
\begin{minipage}{0.5\textwidth}
\resizebox{7.0cm}{4.18cm}{\includegraphics{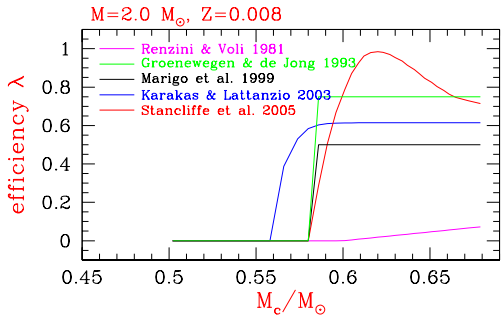}}
\end{minipage}
\hfill
\begin{minipage}{0.5\textwidth}
\caption{Predicted behaviour of the efficiency of the third
dredge-up during the TP-AGB evolution of the same model, according
to various authors}
\end{minipage}
\label{fig_lambda}
\end{figure}

Extensive full AGB calculations have recently provided a more detailed
picture of the third dredge-up and its dependence on stellar mass 
and metallicity, supplying a fitting formalism for two classical quantities: 
the minimum core mass  $M_{\rm c}^{\rm min}(M,Z)$ and the efficiency 
$\lambda(M,Z)$  (Karakas et al. 2002). This allows to include
a more realistic description of the dredge-up process in synthetic 
TP-AGB models (Marigo \& Girardi 2006), 
so as to relax the assumption of constant dredge-up
 parameters characterizing several works in the past literature 
(e.g. Groenewegen \& de Jong 1993, Marigo et al. 1999, 
Mouhcine \& Lan\c{c}on 2003). It is also interesting to see in 
Fig.~\ref{fig_lambda} how
much the predictions for the efficiency $\lambda$ have changed over the years,
e.g. passing from almost no dredge-up in Renzini \& Voli (1981)
to quite extreme values according to Stancliffe et al. (2005).
  
Taking  HBB into account in synthetic TP-AGB models can be done via
two possible ways, namely either analytically or 
with the inclusion of a nuclear
network coupled to envelope integrations. The former scheme has been
adopted in various models (Groenewegen \& de Jong 1993, Mouhcine \&
Lan\c{c}on 2002a, Izzard et al. 2004), but 
its performance is heavily affected by (too) many free parameters.
The best possible approach is clearly the latter one, producing 
results that are closer to those of full AGB calculations
(e.g. Marigo 2001, Gavil\'an et al. 2005).

\section{Stellar yields}
\label{sec_yields}
A direct by-product of AGB evolutionary calculations is the computation
of stellar yields, which are key-ingredients in chemical evolution
models of galaxies. A compilation of available sets of yields from low-
and intermediate-mass stars is presented
in Table~\ref{tab_yields}.

An important consistency requirement to galaxy models is set by
the fuel consumption theorem (Renzini \& Buzzoni 1986) that states
the direct proportionality between the nuclear fuel, $F$, burnt 
during a post main-sequence phase and its contribution 
to the integrated light of a SSP.  
 
Marigo \& Girardi (2001) have shown, in particular, that the nuclear
fuel of the AGB phase can be conveniently split into two contributions, 
one corresponding to the fuel burnt to increase the mass of the C-O core 
and the other related to the fuel ejected from the star in the form
of chemical yields (essentially helium, carbon and oxygen).
In other words, this means  that chemical yields are just a precise fraction
of the stellar emitted light. It follows that 
in chemo-spectro-photometric galaxy models
the basic stellar ingredients, i.e. light and chemical yields,  ought
come from the same set of stellar models or, at least, the reciprocal
consistency of heterogeneous inputs should be verified.
 
\section{Observational constraints to AGB stellar models}
\label{sec_obscons}
Two important constraints will be discussed here, namely: i) 
the counts of AGB stars, and ii) the carbon star luminosity functions (CSLF)
in the Magellanic Clouds (MCs).
Their simultaneous fulfillment should guarantee    
that AGB lifetimes and luminosities are fairly estimated, 
that is the amount AGB fuel in EPS models is consistent with observations.
   
\subsection{AGB stellar lifetimes}
Star counts in MCs' clusters provide quantitative information
on the duration of the AGB phase as a function of stellar mass (and 
metallicity). Further, one can distinguish between M-type and C-type 
lifetimes, $\tau_{\rm M}$ and  $\tau_{\rm C}$ respectively 
(Girardi \& Marigo 2006).
\begin{figure}[!ht]
\plottwo{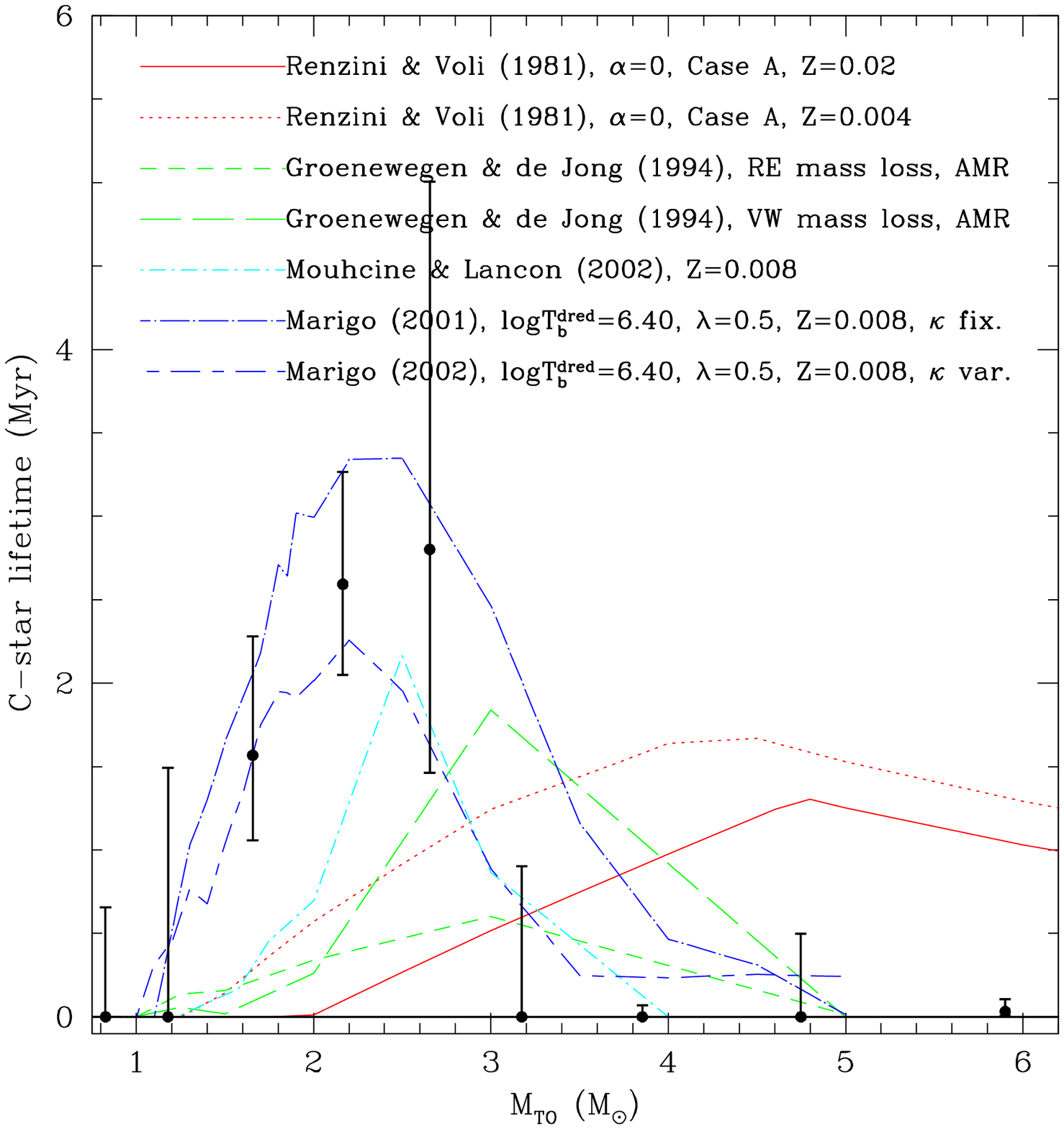}{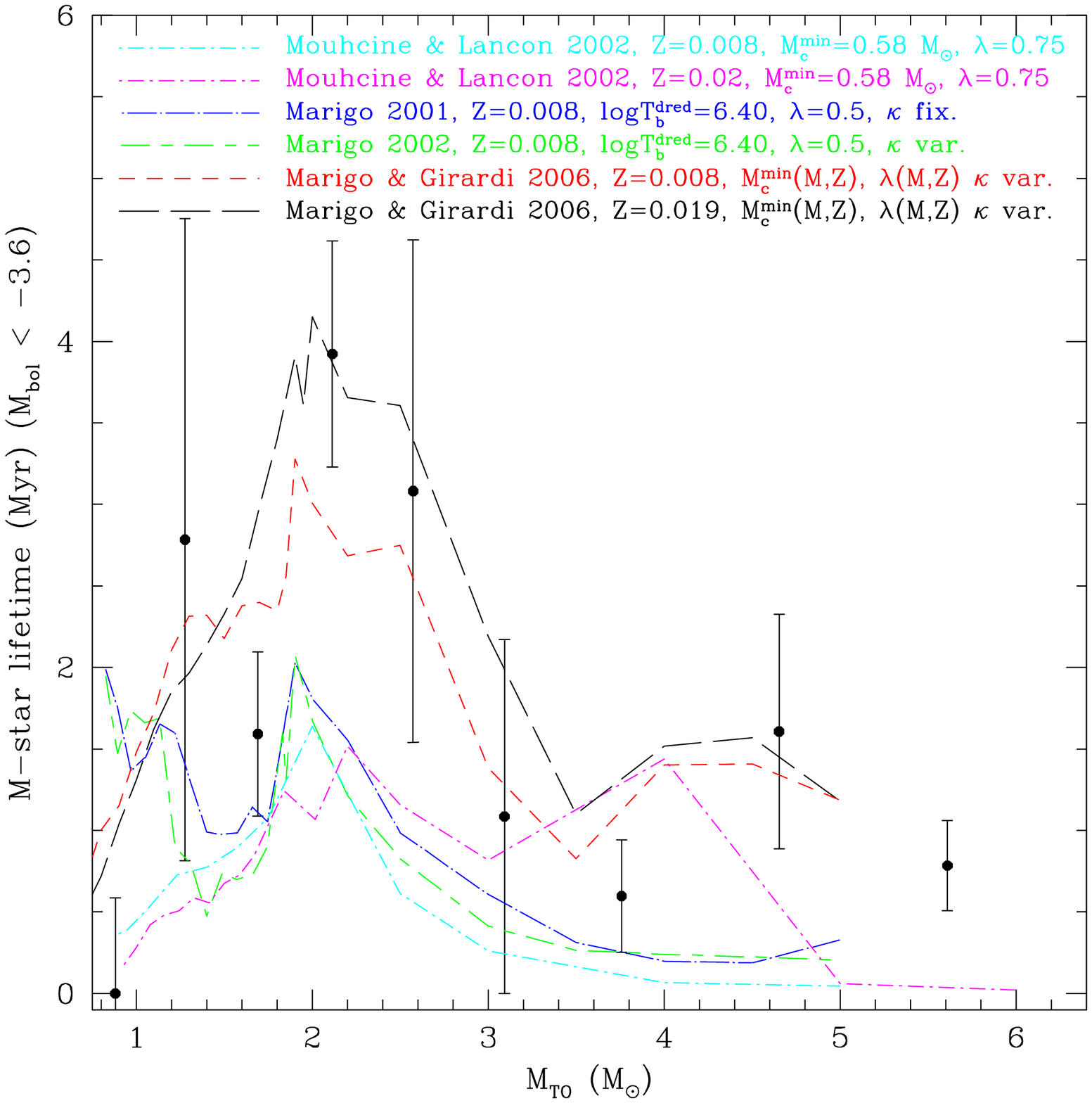}
\caption{left panel: Duration of the C-type (left panel) and M-type
(right panel) stellar phases as a function of the turn-off
stellar mass in LMC clusters (Girardi \& Marigo 2006), compared to  
predictions of various synthetic TP-AGB models. See text for more explanation}
\label{fig_taucm}
\end{figure}
 Figure~\ref{fig_taucm} compares the LMC data with 
the predictions of a few synthetic TP-AGB models, obtained with different
assumptions concerning the third dredge-up, mass-loss rates, molecular
opacities, etc. We note that $\tau_{\rm C}$ varies notably from author
to author, being particularly sensitive to the adopted mass-loss formalism.
Different is the case of $\tau_{\rm M}$, which is practically independent
of mass loss (for AGB stars that later become C-stars), while it
is crucially affected by the third dredge-up, since its duration
is limited by the transition to the C-star domain.
Interestingly, the significant underestimation of  $\tau_{\rm M}$
at $M\approx 2.0-2.5 M_{\odot}$ by models adopting constant 
$M_{\rm c}^{\rm min}$ and $\lambda$ is filled in when accounting
for the dependence of these parameters on mass and metallicity,
as provided by Karakas et al. (2002).  

\subsection{Carbon star luminosity functions}
\label{ssect_cslf}
Observed CSLFs in the MCs pose an important constraint to
AGB models. The long-standing theoretical difficulty to account for
the formation of faint C-stars -- the well-known carbon star mystery designated
by Iben (1981) -- still affects present full AGB
evolutionary calculations (e.g. Izzard et al. 2004; see however the 
results of Stancliffe et al. (2006) for the LMC case). 
Synthetic TP-AGB models indicate that to remove the 
discrepancy one has to invoke an earlier onset (lower $M_{\rm c}^{\rm min}$) 
and higher efficiency (larger $\lambda$) of the
dredge-up in lower-mass models 
(Groenewegen \& de Jong 1993, Marigo et al. 1999). 

\begin{figure}[!ht]
\begin{minipage}{0.49\textwidth}
\resizebox{\hsize}{!}{\includegraphics{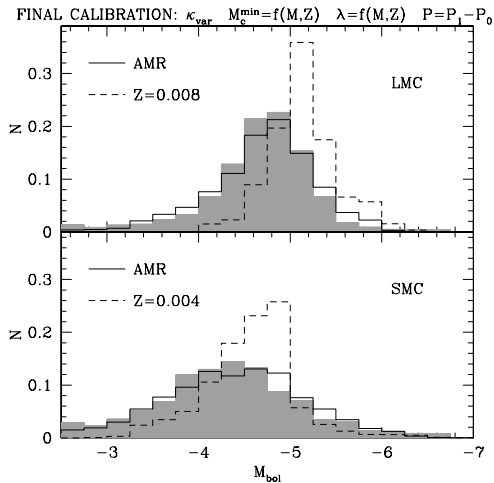}}
\end{minipage}
\hfill
\begin{minipage}{0.49\textwidth}
\caption{Carbon star luminosity functions of the LMC and SMC. Observed
distributions (gray histograms) are compared to  the results of
calibrated synthetic TP-AGB models (Marigo \& Girardi 2006),
assuming either an AMR (continuous line) or a constant metallicity (dashed
line) for the parent galaxies. Models include variable molecular opacities,
and a proper mass- and metallicity-dependence of the third dredge-up based
on Karakas et al. (2002) }
\end{minipage}
\label{fig_cslf}
\end{figure}

However, it should be mentioned that  most simulations of CSLFs have assumed 
a constant metallicity (i.e. $Z=0.008$ for the LMC and 
$Z=0.004$ for the SMC), while  according to 
a recent study (Marigo \& Girardi 2006) this approximation 
should be dropped in favour of a proper coupling between a more realistic
mass- and metallicity-dependent description of the third dredge-up and the
AMR of the parent galaxy (see Fig.~\ref{fig_cslf}). 
By doing so fainter C-stars tend to have 
lower metallicities,  just where the occurrence of the third dredge-up 
is naturally favoured by  AGB models. 
In this way the extent of the
theoretical difficulty turns out to be not so severe as believed so far.  

Recently Guandalini et al. (2006) have suggested that the  carbon star mystery
might be a false problem when accounting for the obscured C-stars, which
would make the  CSLF brighter than previously believed. However, both
observations  (Van loon et al. 2005) and population synthesis models
(see Girardi \& Marigo, this meeting) agree in estimating as 
$\approx 10-20 \%$ the fraction of dust-enshrouded C-stars in the MCs, i.e.
the bulk of the C-star population consists of optically-visible
objects. This implies that the consideration of the obscured C-stars
would not dramatically affect the observed CSLFs, at least in these two
galaxies.

\section{Resolved stellar populations with AGB stars}
\subsection{C-star populations and the C/M ratio}
\begin{figure}[!ht]
\begin{minipage}{0.49\textwidth}
\resizebox{\hsize}{6.6cm}{\includegraphics{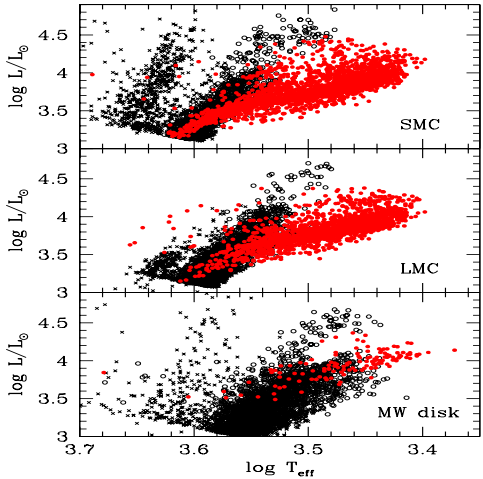}}
\end{minipage}
\hfill
\begin{minipage}{0.49\textwidth}
\caption{HR diagram of three simulated galaxies (Milky Way disk, LMC, and SMC)
including AGB stars (circles), both oxygen-rich (empty circles)
and carbon-rich (filled circles). From Marigo \& Girardi (2006)}
\end{minipage}
\label{fig_hr}
\end{figure}

Recent wide-area near-IR surveys (DENIS, 2MASS) have shown that C-stars
draw a striking red tail in colour-magnitude diagrams, clearly separated
from the branch of oxygen-rich giant stars toward redder colours. 
With the aid of the isochrone synthesis approach,  
Marigo et al. (2003) have shown that the
reproduction of such feature requires the adoption 
in AGB stellar models of molecular
opacities consistently coupled with the surface C/O 
(see Sect.~\ref{ssect_lt}). 
It is interesting to notice in Fig.~\ref{fig_hr} that the C-star
population is expected to be more prominent as we move 
from the Milky Way disk to the SMC, which is essentially driven by
a metallicity effect.

As a matter of fact, whereas the C/M ratio is predicted to be primarily 
a function of metallicity (Mouhcine \& Lan\c{c}on 2003), 
it is expected to change also with population age, hence  
being sensitive to the star formation history 
of the host galaxy (Cioni et al. 2006a). 
Moreover, the mean K-band magnitude of C- and M-type stars 
also varies with the SFH. Cioni et al. (2006ab) take advantage of these 
theoretical predictions in order to map variations 
of mean age and metallicity across the LMC and SMC.

\subsection{Long-period variables}
Despite the huge observational progress achieved thanks to recent 
gravitational  micro-lensing experiments (MACHO, EROS, OGLE), 
few population synthesis studies on pulsating AGB stars 
have been carried out so far (e.g. Groenewegen \& de Jong 1994).
The potentiality of such theoretical analyses is far-reaching. 
For instance, by fitting the observed period distributions 
of OGLE Miras in selected
Bulge fields, Groenewegen \& Blommaert (2005) could infer the presence
of relatively young stellar populations, with ages $\la 3$ Gyr.  

The most intriguing future challenge of EPS models with AGB
stars will be the 
reproduction  
of the  several sequences populated by
variable red giants period-luminosity  diagrams, most of which
are interpreted as due to different pulsation modes (Wood et al. 1999).

\section{Effects of AGB stars on integrated luminosities and colours}
\label{sec_intlc}
One basic prediction of EPS models is the evolution
of integrated luminosities  and broad-band colours (e.g. $B-V$, $V-K$, $J-K$) 
produced by  single stellar populations as a function of age
(e.g. Tantalo et al. 1996, Maraston 1998, Marigo \& Girardi 2001, 
Mouhcine \& Lan\c{c}on 2002a).
The standard benchmark of any theoretical comparison has long been provided
by AGB stars in star clusters in the MCs (Frogel et al. 1990).
Unfortunately, due to the small number statistics of AGB stars in these 
clusters, stochastic fluctuations are of large entity, 
so that MCs' clusters cannot be assimilated  to standard SSP of given age and
metallicity. The best meaningful way to perform a comparison
with observations is to simulate a population of clusters
with the aid of Monte-Carlo techniques (Bruzual \& Charlot 2003). 

\begin{figure}[!ht]
\begin{minipage}{0.49\textwidth}
\resizebox{\hsize}{!}{\includegraphics{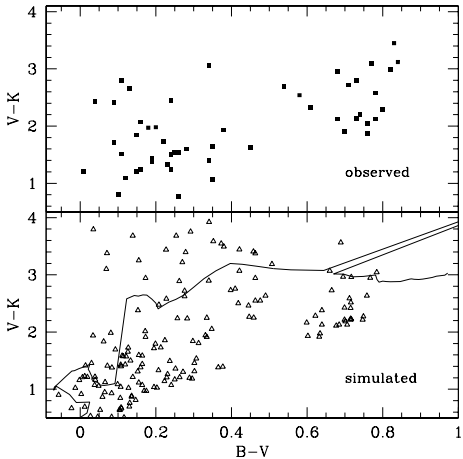}}
\end{minipage}
\hfill
\begin{minipage}{0.49\textwidth}
\caption{Top panel: Observed $(V-K)$  vs $(B-V)$   diagram 
for a sample of LMC clusters of known ages 
(Persson et al. 1988, Bica et al. 1996; Girardi et al. 1995).
Bottom panel: Predictions for a randomly generated 
sample of stellar clusters (empty triangles)
obeying given mass and age distributions. The solid line connects the
values of SSPs of increasing age, i.e. with increasing $(B-V)$. 
The metallicity is assumed
$Z=0.008$ (Marigo \& Girardi 2006) }
\end{minipage}
\label{fig_vkbv}
\end{figure}

For instance, by looking at Fig.~\ref{fig_vkbv},  
we notice that the
observed dispersion of the data (top panel) is recovered with a  
synthetic sample  of clusters (bottom panel), 
while limiting the comparison to a sequence of SSPs 
of varying age (solid line) may be even misleading. 
 
AGB stars are important contributors 
to their integrated bolometric and near-IR luminosities with a maximum  
located at an age of $\approx 1$ Gyr, accounting for about $40-80 \%$
of the total cluster's luminosity (see  Frogel et al. 1990; Maraston 2005).
Interestingly, the position of the maximum reflects the peak
in TP-AGB lifetimes at $M\approx 2.0-2.5 M_{\odot}$, as shown in 
Fig.~\ref{fig_taucm}.

Going from younger to older ages the first appearance of more massive 
AGB stars at $\approx 10^{8}$ Gyr produces a significant 
increase of colours like $V-K$  
(e.g. Maraston 1998), with a rising slope that depends much on the
details of processes like mass loss  and HBB (Girardi \& Bertelli 1998).  
A further contribution to attain redder colours (e.g. $V-K$, $J-K$) is provided
by C-stars, with a maximum effect at ages $\sim 0.5-1.0 \times 10^{9}$ yr
 corresponding to  turn-off masses of $\sim 2.0-2.5 M_{\odot}$ 
(Mouhcine \& Lan\c{c}on 2002a).  

As first shown by 
Bressan et al. (1998, see also Mouhcine 2002, Piovan et al. 2003) 
accounting for the effect of dusty envelope around AGB stars 
has a strong impact on the predicted spectral properties of galaxies, e.g. 
increasing the integrated light emission in the mid-infrared 
of an intermediate-age population by one order of magnitude.

A potentially powerful application of such EPS models 
including  dust emission from AGB stars is the possibility
to break the age-metallicity degeneracy that affect early-type galaxies,
with the aid of a suitable combination of optical$+$mid-IR$+$near-IR pass-bands.
A recent result to be mentioned is 
the detection of a broad silicate emission at $10 \mu$m in the
Spitzer IRS spectra of a large sample of early-type galaxies in Virgo, which
appears to be consistent with the contribution from an underlying 
population of unresolved dusty oxygen-rich AGB stars (Bressan et al. 2006).

\acknowledgements 
This work is funded by the University of Padova (Progetto di ricerca
di Ateneo CPDA052212).


\end{document}